\documentclass[12pt]{article}

\usepackage{amsmath,amssymb}
\usepackage{amsthm}
\usepackage[top=2.828cm,bottom=2.828cm, left=3cm,right=3cm]{geometry}
\usepackage{bm}
\usepackage{graphicx}
\usepackage{xcolor}
\numberwithin{equation}{section}
\usepackage{cite}
\usepackage{ascmac}
\usepackage[breaklinks=true,linktocpage=true,pagebackref=false]{hyperref}
\usepackage{mathrsfs}
\newcommand{\diag}{\,{\rm diag}\,}
\newcommand{\link}{\,{\rm link}\,}
\newcommand{\er}[1]{Eq.~\eqref{#1}}
\newcommand{\ers}[1]{Eqs.~\eqref{#1}}

\newcommand{\vev}[1]{\left\langle #1 \right\rangle}

\newcommand{\tr}{\text{tr}}

\newcommand{\bb}{\mathbb}

\newcommand{\hph}{\hphantom}
\renewcommand{\b}{\bar}

\newcommand{\sr}{\sqrt}
\newcommand{\bs}{\boldsymbol}
\newcommand{\df}{\dfrac}
\newcommand{\fr}{\frac}

\newcommand{\der}{\partial}

\renewcommand{\(}{\left(}
\renewcommand{\)}{\right)}

\newcommand{\wed}{\wedge}
\newcommand{\bmx}{\left(\begin{matrix}}
\newcommand{\emx}{\end{matrix}\right)}
\newcommand{\mtx}[1]{\bmx #1 \emx}

\begin{document}
\begin{titlepage}
\hfill RIKEN-QHP-409, KEK-TH-2180
\vspace{-1em}
\def\thefootnote{\fnsymbol{footnote}}%
   \def\@makefnmark{\hbox
       to\z@{$\m@th^{\@thefnmark}$\hss}}%
 \vspace{3em}
 \begin{center}%
  {\Large 
Topological order in a color-flavor locked phase 
\\
of $(3+1)$-dimensional $U(N)$ gauge-Higgs system
\par
   }%
 \vspace{1.5em}
  {
Yoshimasa Hidaka${}^{1,2}$\footnote{hidaka@riken.jp},
Yuji Hirono${}^{3,4}$\footnote{yuji.hirono@apctp.org},
Muneto Nitta${}^5$\footnote{nitta@phys-h.keio.ac.jp},
\\
Yuya Tanizaki${}^6$\footnote{ytaniza@ncsu.edu},
and
Ryo Yokokura${}^{7,5}$\footnote{ryokokur@post.kek.jp}
   \par
   }%
 \vspace{1.5em}
{\small \it
${}^1$Nishina Center, RIKEN, Wako 351-0198, Japan
\\
${}^2$RIKEN iTHEMS, RIKEN, Wako 351-0198, Japan
\\
${}^3$Asia Pacific Center for Theoretical Physics, Pohang 37673, Korea 
\\
${}^4$Department of Physics, POSTECH, Pohang 37673, Korea
\\
${}^5$Department of Physics \& 
Research and Education Center for Natural Sciences,\\
Keio University, Hiyoshi 4-1-1, Yokohama, Kanagawa 223-8521, Japan
\\
${}^6$Department of Physics, North Carolina State University, Raleigh, NC 27607, USA
\\
${}^7$KEK Theory Center, Tsukuba 305-0801, Japan
 \par
}
  \vspace{1em} 
 \end{center}%
 \par
\begin{abstract}
We study a $(3+1)$-dimensional $U(N)$ gauge theory with $N$-flavor fundamental scalar fields, whose color-flavor locked (CFL) phase has topologically stable non-Abelian vortices. 
The $U(1)$ charge of the scalar fields must be $Nk+1$ for some integer $k$ in order for them to be in the representation of $U(N)$ gauge group. 
This theory has a $\mathbb{Z}_{Nk+1}$ one-form symmetry, and it is spontaneously broken in the CFL phase, i.e., the CFL phase is topologically ordered if $k\not=0$. 
We also find that the world sheet of topologically stable vortices in CFL phase can generate this one-form symmetry.
\end{abstract}
\end{titlepage}
 \setcounter{footnote}{0}%
\def\thefootnote{$*$\arabic{footnote}}%
   \def\@makefnmark{\hbox
       to\z@{$\m@th^{\@thefnmark}$\hss}}%
\tableofcontents
\section{Introduction}
Order or disorder is a basic concept to classify classical and quantum phases of matter in modern physics. 
Classically, phases such as liquid and solid are characterized by local order parameters according to Landau's symmetry breaking theory. 
Local order parameters, however, are insufficient to classify quantum phases; 
topological order does not have local order parameters,
but still they lead to distinct phases. 
Topologically ordered states 
exhibit exotic properties such as fractional statistics, topological degeneracy of ground states, and long range entanglement~\cite{Wen:1989zg,Wen:1989iv,Wen:1990zza,Wen:1991rp,Wen:1995qn,Hansson:2004wca,Kitaev:2005dm,Levin:2006zz,Chen:2010gda}. 
According to the recent developments in the understanding of the symmetry classification of quantum phases, some of the topologically ordered phases can be classified by a spontaneous breakdown of a higher form symmetry~\cite{Banks:2010zn,Kapustin:2014gua,Gaiotto:2014kfa},  
which is a symmetry acting on extended objects such as vortex lines and domain walls~\cite{Nussinov:2009zz}.
In addition to ordinary symmetries, higher form symmetries can be employed to classify quantum phases. 

Fractional quantum Hall system and a toric code~\cite{Tsui:1982yy,Laughlin:1983fy,Kitaev:1997wr} are typical examples of $(2+1)$-dimensional topological orders, and they possesses spontaneously-broken one-form symmetries.  
Low-energy effective theories of those can be expressed as Chern-Simons or $BF$-type topological gauge theories~\cite{Horowitz:1989ng,Allen:1990gb,Blau:1989dh} (see Refs.~\cite{Birmingham:1991ty} for review). 
In the case of fractional quantum Hall systems,  
the charged object and symmetry generator are both Wilson loops. An anyon is attached to the endpoint of an open Wilson line while the trajectory can be represented by the Wilson line; a braiding of trajectories of two anyons results in a fractional phase when the Wilson lines are linked. 
A similar situation occurs in $(3+1)$ dimensions. 
An example of a topological order in $(3+1)$ dimensions is provided by s-wave BCS superconductors~\cite{Hansson:2004wca}. 
In the superconducting phase, $\mathbb{Z}_2$ one-form (and also two-form) symmetry emerges at low energies below Cooper-pair binding energy. 
Objects charged under these symmetries are a Wilson loop and a surface operator, respectively. 
In the superconducting phase, the Wilson loop exhibits a perimeter law, i.e., $\bb{Z}_2$ one-form symmetry is spontaneously broken.
There are string-like excitations called Abrikosov-Nielsen-Olesen (ANO)  vortices~\cite{Abrikosov:1956sx,Nielsen:1973cs}, 
and their world sheets can be regarded as generators of the one-form symmetry.
Unlike the $(2+1)$-dimensional topological order, a braiding phase appears between a particle and a vortex.

Non-Abelian gauge theories admit a non-Abelian generalization of ANO vortices in the Higgs phase. 
For example, 
$U(N)$ gauge theory coupled with $N \times N$ complex scalar fields 
in the fundamental representation 
admits non-Abelian vortices in the color-flavor locked (CFL) phase
\cite{Hanany:2003hp,Auzzi:2003fs,Eto:2004rz,Eto:2005yh,Gorsky:2004ad,Eto:2006cx,Eto:2006db}
\footnote{
Although those findings were made in supersymmetric models, 
the supersymmetry is not essential for the existence of non-Abelian vortices. 
}
(see Refs.~\cite{Tong:2005un,Eto:2006pg,Shifman:2007ce,Shifman:2009zz} for a review).
Those vortices are accompanied by $\mathbb{C}P^{N-1}$ 
Nambu-Goldstone modes
which are localized along them. 
QCD at high densities is also in the CFL 
phase~\cite{Alford:1997zt,Alford:1998mk} 
(see Refs.~\cite{Ren:2004nn,Stephanov:2004wx,Alford:2007xm,Fukushima:2010bq}
for a review), 
and it admits similar non-Abelian vortices 
\cite{Balachandran:2005ev,Nakano:2007dr,Eto:2009kg} 
accompanied by $\mathbb{C}P^2$ moduli 
\cite{Nakano:2007dr,Eto:2009bh,Eto:2009tr} 
(see Ref.~\cite{Eto:2013hoa} for a review).
One natural question is whether these theories are topologically ordered or not, 
since they can be regarded as non-Abelian extensions of superconductors \cite{Hansson:2004wca}.
In the case of the CFL phase of dense QCD, 
this question was first addressed in Ref.~\cite{Cherman:2018jir}
and later elaborated on in Ref.~\cite{Hirono:2018fjr}.
It turned out that the CFL phase of QCD is {\it not} topologically ordered. 
This is because the emergent discrete two-form symmetry is unbroken due to the interaction between vortices and massless Nambu-Goldstone bosons. 
Those particles mediate the forces between vortices, 
resulting in a log-confining potential between a vortex and an antivortex. 
This implies the vanishing of the expectation value of the vortex surface operator at a large surface, which is the order parameter for this symmetry. 

In this paper, we discuss a possible appearance of topological order in quantum field theories that have non-Abelian vortices. 
We study a $U(N)$ gauge theory coupled to 
complex scalar fields in the fundamental representation of 
$SU(N)$ gauge symmetry as well as 
$SU(N)/\bb{Z}_N$ flavor symmetry. 
In order for the scalar fields to be in the representation of $U(N)$, 
their $U(1)$ charge must be taken as $Nk+1$ with some integer $k$. 
Unlike the CFL phase of QCD~\cite{Hirono:2018fjr}, 
we find that 
the CFL phase in the $U(N)$ gauge theories is topologically ordered 
if $k \neq 0$,  
while the previously considered $U(N)$ theories with $k = 0$ 
\cite{Tong:2005un,Eto:2006pg,Shifman:2007ce,Shifman:2009zz}
are not. 
This is because the system has a 
$\bb{Z}_{Nk+1}$ one-form symmetry, and it is spontaneously broken in the Higgs phase, which means that this phase has a topological order. 
We also find that the $\bb{Z}_{Nk+1}$ two-form symmetry emerges at low energies, and it is also spontaneously broken. 

This paper is organized as follows.
In Sec.~\ref{AH}, we review the topological order in 
the $(3+1)$-dimensional Abelian Higgs model.
In Sec.~\ref{UN}, we discuss the existence of topologically 
ordered phase
in $U(N) $ gauge theories with $N$-flavor scalar fields.
Section \ref{sum} is devoted to a summary and discussions. 
We summarize some properties of the delta function forms and linking numbers in Appendix~\ref{bflink}.
In this paper, we use the Euclidian metric, 
$\delta_{mn} =\diag(+1,+1,+1,+1)$,
where $m,n,...$ are indices of the spacetime coordinates.

\section{Topological order in $U(1)$ gauge theory}
\label{AH}
We here review the appearance of topological order 
in the low-energy effective theory of the Abelian Higgs
model in $(3+1)$ dimensions~\cite{Hansson:2004wca}. 
We derive a dual theory of the Abelian
Higgs model with a charge $k$ scalar field.  
The derived theory is the so-called 
$BF$-theory~\cite{Horowitz:1989km,Blau:1989bq} at level $k$. 
We then show that there is an emergent
$\bb{Z}_k$ two-form symmetry in addition to 
the $\bb{Z}_k $ one-form symmetry in the original action, 
and both of the $\bb{Z}_k$ symmetries are spontaneously broken, 
by calculating correlation functions of Wilson loops and surface 
operators~\cite{Horowitz:1989km,Blau:1989bq,Oda:1989tq,Chen:2015gma}.

\subsection{Dual $BF$-theory from Abelian Higgs model}\label{sec:dual_BF_theory}
Here, we derive the $BF$-theory via an Abelian duality.
We begin with the low-energy theory of the Abelian Higgs model 
in $(3+1)$ dimensions described by the action, 
\begin{equation}
 {S}_{\rm AH} = 
\int \fr{\xi}{2}
|d\chi - k A|^2
+
\fr{1}{2 e^2} \int |dA|^2 , 
\label{190122.2305} 
\end{equation}
where $\chi$ is a $2\pi$-periodic scalar field, 
$A$ is a $U(1)$ gauge field, $d$ is an exterior derivative operator,
$*$ is a Hodge's star operator, 
$\xi$ is a parameter with mass-dimension $2$, $k$ is an integer,
and $e$ is a coupling constant.
The symbol $\int |\omega|^2$ for a $p$-form field $\omega$ denotes
\begin{equation}
 \int |\omega|^2 = \int \omega \wed * \omega 
= \int d^4x \sr{g} \fr{1}{p!}\omega_{m_1...m_p} \omega^{m_1...m_p},
\end{equation}
where $g = \det (g_{mn})$.
The scalar field $\chi$ and the parameter $\sr{\xi/2}$ 
can be understood as the phase component and 
the vacuum expectation value of the amplitude of the Higgs field, respectively.
Photons are massive via the Higgs mechanism. 
The action has a $U(1)$ gauge symmetry 
$\chi \to \chi+k\lambda^{(0)}$ and $A\to A+d\lambda^{(0)}$,
where $\lambda^{(0)}$ is a zero-form gauge parameter.
In addition, the action has a $\bb{Z}_k$ one-form global symmetry
given by $A \to A + \fr{n}{k}\epsilon^{(1)}$ with
the condition $\int_{\cal C} \epsilon^{(1)}\in 2\pi \bb{Z}$ 
for a closed loop ${\cal C}$ and $n \in \bb{Z}$,
since the charge of the matter field $\chi$ is $k$.

The action (\ref{190122.2305}) can be dualized to 
a system with a two-form gauge field as follows.
We introduce the following first order action:
\begin{equation}
 {S}_{\rm AH,1st} = 
\fr{1}{8 \pi^2{\xi}}
\int |H|^2
+
\fr{1}{2 e^2}\int |dA|^2
-
\fr{i }{2\pi}\int H \wed (d\chi - kA),
\label{190122.2309} 
\end{equation}
where $H$ is a three-form field.
The equation of motion for $H$ gives us the original action in \er{190122.2305}.
Instead, the equation of motion of  $\chi$ leads to $dH = 0$
and thus the three-form field can be written as
\begin{equation}
 H = dB,
\label{190122.2316}
\end{equation}
where  $B$ is a two-form $U(1)$ gauge field\footnote{To be more precise, $H\in H^3(X,2\pi\mathbb{Z})$, where $X$ is the spacetime manifold.}. 
The one-form gauge transformation is $B \to B + d\lambda^{(1)}$, 
where $\lambda^{(1)}$ is a one-form gauge parameter.
Substituting the solution in \er{190122.2316} into 
the first order action in \er{190122.2309},
 we obtain the dual action:
\begin{equation}
 {S}_{\rm AH,dual} = 
\fr{1}{8\pi^2{\xi}}\int 
|dB|^2
+ \fr{1}{2 e^2} \int
|dA|^2 
-\fr{i k}{2\pi}\int B \wed  dA.
\label{190122.2317}  
\end{equation}
In the presence of the topological coupling $B \wed  dA$, 
both of the one-form and two-form gauge fields become massive.
Therefore, the low-energy effective action, where we can neglect the kinetic term of $A$ and $B$, becomes 
the $BF$-action: 
\begin{equation}
 {S}_{BF} 
= 
-\fr{ik}{2\pi} \int B \wed  dA.
\label{190122.2323}  
\end{equation}
The gauge fields satisfy the usual Dirac quantization condition, 
\begin{equation}
\int_{\mathcal S} d A \in 2\pi \mathbb Z, 
\quad 
\int_{\mathcal V} d B \in 2\pi \mathbb Z, 
\end{equation}
where $\mathcal S$ and $\mathcal V$ are closed 2- and 
3-dimensional manifold, respectively.

\subsection{$BF$-theory and topologically ordered phase}

The $BF$-theory describes topologically ordered states. 
We show that there is an emergent $\bb{Z}_k$ two-form symmetry 
in addition to 
the $\bb{Z}_k$ one-form symmetry, and both of them are broken 
spontaneously.

The action \eqref{190122.2323} is invariant under 
one-form and two-form gauge transformations:
\begin{equation}
\begin{split}
A\to A+d\lambda^{(0)},\qquad
B\to B+d\lambda^{(1)},
\end{split}
\end{equation}
where $\lambda^{(0)}$ and $\lambda^{(1)}$ represent zero- and one-form gauge parameters.
In addition, the action has symmetries under global one- and two-form transformations:
\begin{equation}
\begin{split}
A\to A+\frac{n}{k}\epsilon^{(1)},\qquad
B\to B+\frac{n}{k}\epsilon^{(2)},
\end{split}
\end{equation}
where $n\in \mathbb{Z}$, $d\epsilon^{(1)}=0$ and $d\epsilon^{(2)}=0$. They are properly normalized as
\begin{equation}
\begin{split}
\int_{\cal C} \epsilon^{(1)} \in 2\pi \mathbb{Z},\qquad
\int_{\cal S} \epsilon^{(2)} \in 2\pi \mathbb{Z}.
\end{split}
\label{190227.2054}
\end{equation}
The charged object and  the symmetry generator 
of the one-form symmetry
are 
a Wilson loop on a closed path ${\cal C}$ and 
a surface operator on a closed surface ${\cal S}$~\cite{Gukov:2006jk,Gukov:2008sn},
\begin{equation}
\begin{split}
W({\cal C}) = e^{i\int_{{\cal C}} A}, 
\qquad V({\cal S}) = e^{i\int_{{\cal S}} B},
\end{split}
\end{equation}
respectively.  
On the other hand, 
as for the two-form symmetry, 
$V({\cal S})$ is the charged object and $W({\cal C})$ is the symmetry generator.
Indeed, one can readily find that $W(\mathcal{C})$ and $V(\mathcal{S})$ are topological, i.e., they do not depend on the small change of $\mathcal{C}$ and $\mathcal{S}$ thanks to the equation of motion $d A=0$ and $d B=0$, and this is nothing but the conservation law~\cite{Gaiotto:2014kfa}. 

The topological nature of symmetry generators implies that the expectation value of $V({\cal S})$ on the  spacetime manifold $\mathbb{R}^{4}$ is trivial
because the symmetry generator can shrink to the point:
\begin{equation}
\begin{split}
\langle V({\cal S})\rangle =\langle 1\rangle =1,
\end{split}
\label{190302.2148}
\end{equation}
where  the expectation value of an object $\mathcal{O}$ is given as
\begin{equation}
\begin{split}
\langle\mathcal{O}\rangle =\mathcal{N}\int \mathcal{D}B \mathcal{D}A e^{-S_{BF}}\mathcal{O}.
\end{split}
\end{equation}
Here the normalization factor $\mathcal{N}^{-1}\equiv \int \mathcal{D}B \mathcal{D}A \exp(-S_{BF})$ is chosen such that $\langle 1\rangle =1$.
Similarly, one can find $\langle W({\cal C})\rangle=1$.

In contrast, the correlation function of the charged object $W({\cal C})$ and the symmetry generator $V({\cal S})$ is nontrivial when they are linked:
\begin{equation}
\vev{V({\cal S})W({\cal C}) } =e^{i\phi}\langle W({\cal C})\rangle,
\label{190302.2001}
\end{equation}
with $\phi=-2\pi  \link({\cal S},{\cal C})/{k}$.
Here $ \link({\cal S},{\cal C})$ denotes the linking number between  ${\cal S}$ and ${\cal C}$.
This relation $V({\cal S})W({\cal C})  =e^{i\phi} W({\cal C})$ is nothing but the transformation of the Wilson loop under the one-form symmetry~\cite{Gaiotto:2014kfa}
(For the detailed derivation of these relations in the path integral formulation, see Appendix~\ref{bflink}).
As with the case of ordinary symmetries, the nonvanishing expectation value of the charged object is the signal of symmetry breaking\footnote{
More precisely, the nonvanishing expectation value of charged object $W({\cal C})$ ($V({\cal S})$) with the large length (area) limit is the signal of spontaneous breaking of the one- (two-) form 
symmetry~\cite{Gaiotto:2014kfa,Lake:2018dqm}. 
Since $W({\cal C})$ and ${V}({\cal S})$ are topological in the $BF$-theory, their expectation values are independent of the choice of ${\cal C}$ and ${\cal S}$.
}. Since both $\vev{W({\cal C})} $ and $\vev{V({\cal S})}$ are nonvanishing,  both one- and two-form symmetries are spontaneously broken.

As is seen in the following, the link between symmetry generators leads to the important properties of topological order such as degeneracy of ground state depending on a spatial manifold, and braiding statistics.

\begin{figure}
\begin{center}
\includegraphics[width=0.7\linewidth]{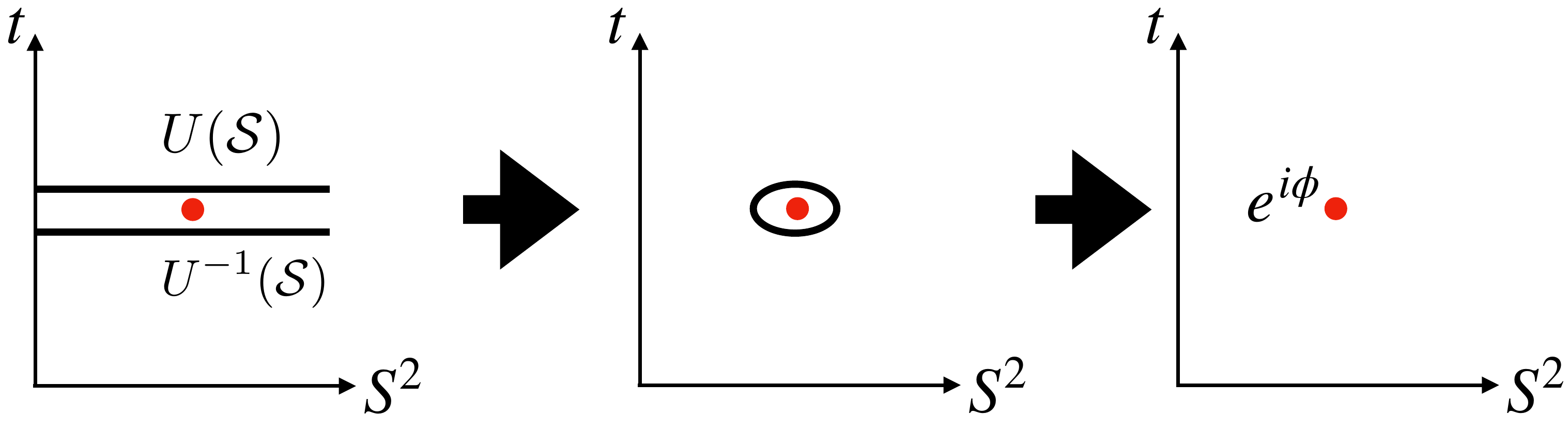}
\caption{Graphical representation of Eq.~\eqref{eq:intersection}:$V({\cal S}) W({\cal C}) V^{-1}({\cal S}) = e^{i\phi} W({\cal C})$.}
\label{fig:commutationRelation}
\end{center}
\end{figure}

\paragraph{Ground state degeneracy}
\begin{figure}
\begin{center}
\includegraphics[width=0.6\linewidth]{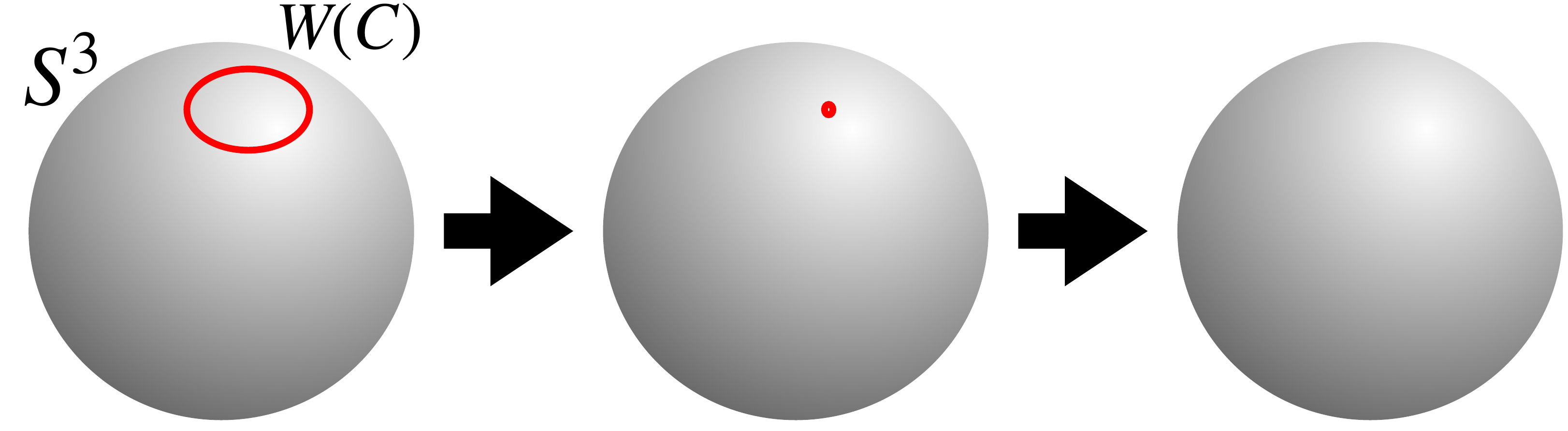}
\caption{Graphical representation of
$W({\cal C})| \Omega \rangle =| \Omega\rangle$ on $S^{3}$.}
\label{fig:deformation}
\end{center}
\end{figure}
One of the key properties of a topological order state is 
the ground state degeneracy depending on the topology of spatial manifold. 
This can be understood as a consequence of spontaneous breaking of higher form symmetries.
If the spatial manifold $\mathcal{M}$ is trivial, e.g., $\mathcal{M}=S^{3}$ or $\mathbb{R}^{3}$,  the generator of higher form symmetry cannot nontrivially act on the vacuum.
This is because the symmetry generator can deform to the point on $S^{3}$ and it vanishes (See Fig.~\ref{fig:deformation}). 
In this case, there is no degeneracy associated with the spontaneous breaking of higher form symmetries. 
In contrast, when $\mathcal{M}$ is nontrivial, more precisely, both $\pi_{1}(\mathcal{M})$ and $\pi_{2}(\mathcal{M})$ are nontrivial,
the surface and line operators can act on the vacuum.
As an example, we consider the manifold  $\mathcal{M}=S^{2}\times S^{1}$,  and choose the surface and line operators as
$V({\cal S}) = \exp{i\int_{{\cal S}} B}$ and  $W({\cal C}) = \exp{i\int_{{\cal C}} A}$ with ${\cal S}=S^{2}$ and ${\cal C}=S^{1}$. On this manifold, these operators satisfy the following relation
in the operator formalism at equal time:
\begin{equation}
\begin{split}
V({\cal S}) W({\cal C}) V^{-1}({\cal S}) = e^{i\phi} W({\cal C}),
\label{eq:intersection}
\end{split}
\end{equation}
with $\phi=2\pi/k$. The graphical representation is shown in Fig.~\ref{fig:commutationRelation}.
Equation~\eqref{eq:intersection} implies the degeneracy of ground state, 
which can be shown as follows:
Since the unitary operator $V({\cal S})$ is a symmetry generator, we can choose a vacuum $|\Omega\rangle$ as the eigenstate of $V({\cal S})$ with the eigenvalue $e^{i\theta}$.
$W({\cal C})$ is also a symmetry generator, so that the state $|\Omega'\rangle:=W({\cal C})|\Omega\rangle$ has the same energy as $|\Omega\rangle$.
If Eq.~\eqref{eq:intersection} is satisfied, $|\Omega\rangle$ and $|\Omega'\rangle$ must be different vacua.
To see this, let us consider the overlap of vacua $\langle \Omega| \Omega'\rangle= \langle \Omega| V|\Omega\rangle $.
Using Eq.~\eqref{eq:intersection}, we find
\begin{equation}
\begin{split}
\langle \Omega| \Omega'\rangle& =  e^{-i\phi} \langle \Omega| VW V^{-1} |\Omega\rangle\\
& =  e^{-i\phi} \langle \Omega| e^{-i\theta}V e^{i\theta}|\Omega\rangle\\
& =  e^{-i\phi} \langle \Omega|\Omega'\rangle,
\end{split}
\end{equation}
where we have used the fact that
 $|\Omega\rangle$ is the eigenstate of $V$.
Since $\phi=2\pi/k\neq0$, the vacua must perpendicular to each other,  $\langle \Omega|\Omega'\rangle=0$. 
That is, the vacuum is degenerate.
More specifically, the vacuum is $k$-fold degenerate on the spatial manifold $\mathcal{M}=S^{2}\times S^{1}$.

\begin{figure}
\begin{center}
\includegraphics[width=0.5\linewidth]{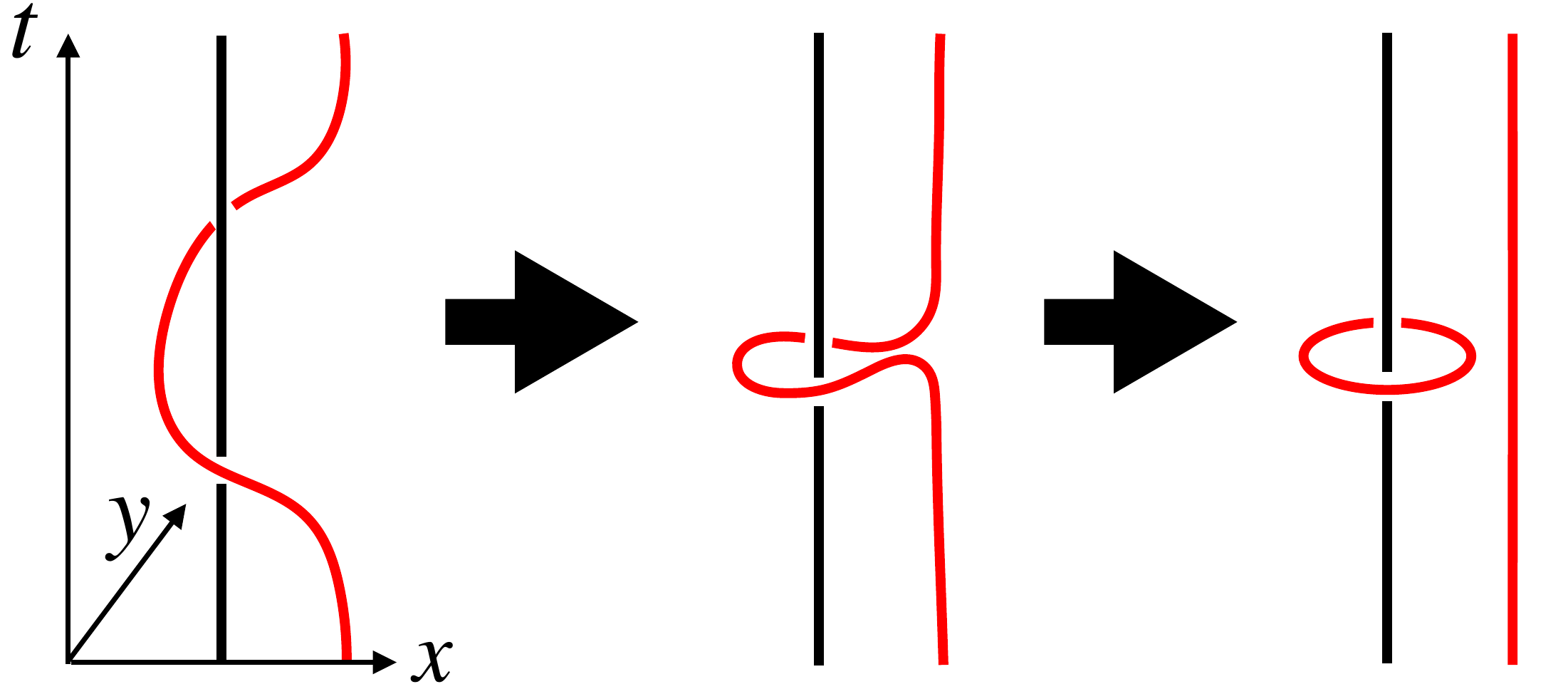}
\caption{World sheet of a vortex (black line) and a world line of a particle (red line). }
\label{fig:braiding}
\end{center}
\end{figure}

\paragraph{Braiding phases}
Another property of a topologically ordered state is the existence of anyonic braiding phases: when two particles are exchanged, the quantum state acquires a phase. When the phase is not $\pm1$ for identical particles, they are anyons. 
In $(3+1)$ dimensions, there is a braiding phase between a particle and a vortex.
For an open line operator, a particle (point) operator can be attached to the boundary of the line operator.
The particle operator is not arbitrary, but it needs to respect the gauge symmetry. 
Similarly, a vortex operator can be attached to the boundary of an open surface operator. 
The trajectories of the particle and vortex are represented as the world line and world sheet, respectively.
The left figure in Fig.~\ref{fig:braiding}  shows their braiding trajectory. 
Since the surface and line operators are topological, it can be deformed into the right figure in Fig.~\ref{fig:braiding}. 
This trajectory causes the phase $2\pi/k$ relative to the straight trajectory.
The half of  the linking phase can be understood as the exchanging phase of the particle and vortex.

\section{$U(N)$ gauge theory in the color-flavor locked 
phase}\label{UN}

In this section, we discuss a $(3+1)$-dimensional $U(N)$ gauge theory coupled to scalar fields. 
The CFL phase of this theory has non-Abelian vortices are topologically stable excitations. 
We show that a topological order appears in the CFL phase and discuss fractional braiding statistics between non-Abelian vortices and quasiparticles.

\subsection{Non-Abelian vortices in color-flavor locked phase}

Here let us introduce our model. 
We consider a $U(N)_{\rm c} $ gauge theory 
coupled with $N$-flavor scalar fields $\phi_f$, with $f=1,\ldots, N$. 
Each $\phi_f$ is the $\mathbb{C}^N$-valued scalar field, and its representation of the gauge group is taken as 
\begin{equation}
R_k(g) \phi=\det(g)^k g\cdot \phi
\label{190310.1918}
\end{equation}
for $g\in U(N)_c$ with some integer $k$. 
We denote $U(N)_{\rm c}$ gauge fields as $A$, then the field strength is given as 
\begin{equation}
F=d A +i A\wedge A. 
\end{equation}
The Lagrangian of the model we consider is given by 
\begin{equation}
S_{\rm NA}
 = \fr{1}{2g_1^2} \int  \tr(F \wed *F) 
+ \fr{1}{2g_2^2} \int \tr(F)\wed *\tr(F)
+\int  D\b\phi_{f} \wed *D \phi_{f}
 + \int  V(\phi,\b\phi) *1. 
\label{190117.0200}
\end{equation}
Here, $g_1$ and $g_2$ 
are gauge coupling constants. 
Covariant derivatives on the scalar fields are given by 
the transformation law of $\phi_{f}$ in \er{190310.1918} 
as
\begin{equation}
D_m \phi_{f} 
= (\der_m -i k \tr[A_m]{\bf 1}_N-i A_m)\phi_f.
\end{equation}
The potential $V$ shall be chosen so that the theory is in a deep Higgs regime, but the details are not important for our discussion.

The flavor symmetry of this theory is 
$SU(N)_{\rm f}/(\mathbb{Z}_N)_{\rm f}$.
Note that the center of $SU(N)_{\rm f}$, $(\bb{Z}_N)_{\rm f}$, 
is absorbed into the 
gauge group $U(N)_{\rm c}$.
In addition to this ordinary symmetry, this theory has the one-form symmetry. 
We consider the transformation on transition functions $g\to g e^{i \alpha}$, and then the representation matrix $R_k(g)$ changes as 
\begin{equation}
R_k(g)\to e^{i(Nk+1)\alpha}R_k(g). 
\end{equation}
When $\alpha$ is quantized to $2\pi/(Nk+1)$, the dynamical fields are not affected, but the $U(N)$ Wilson loop can detect this phase $\alpha$. This means that the theory has $\mathbb{Z}_{Nk+1}$ one-form symmetry. 
In the following, we set 
\begin{equation}
q=Nk+1. 
\end{equation}

Now, let us consider the vacuum structure. 
The minimum of the potential is realized by 
\begin{equation}
 \vev{
 \b \phi^{c f_1}
 \phi_{cf_2}} = \frac{\xi}{2} {\bf 1}_N . 
\end{equation}
where $\bs{1}_N$ is the $N$-dimensional unit matrix 
in the flavor space,
and the subscript or superscript $c = 1,...,N$ denote the index of the fundamental or antifundamental 
representations of $U(N)_{\rm c}$, respectively. 
Therefore, the flavor symmetry $SU(N)_f/(\mathbb{Z}_N)_{\rm f}$ 
is unbroken, but the one-form symmetry is spontaneously broken:
\begin{equation}
\mathbb{Z}_q^{(\mbox{one-form})}\to 1.
\end{equation}
At the mean-field level, 
this is realized by fixing the gauge so that 
\begin{equation}
( \vev{\phi_{cf}} )=
\sr{\fr{\xi}{2}} {\bf 1}_N, 
\end{equation}
As a result, all the gauge fields are Higgsed,
and there is no massless Nambu-Goldstone mode. 
The symmetry breaking pattern is given by 
\begin{equation}
\frac{U(N)_{\rm c} \times SU(N)_{\rm f} }
{
(\mathbb Z_N)_{\rm f}
}
 \to 
\mathbb{Z}_q\times  \frac{
 SU(N)_{\rm c+f}  
}{
(\mathbb Z_N)_{\rm f}
}
.
\end{equation}
The vacuum expectation value is invariant under 
the simultaneous rotations of color and flavor, 
$SU(N)_{{\rm c}+ {\rm f}}$. 
That is why it is called the CFL phase. 
This phase admits topological vortices. 
The vacuum manifold is given by 
\begin{equation}
 \df{
\df{U(N)_{\rm c} \times SU(N)_{\rm f}}{(\bb{Z}_N)_{\rm f}}
 }
 {
\bb{Z}_q \times \df{SU(N)_{{\rm c}+{\rm f}} }{(\bb{Z}_N)_{\rm f}}
 }
\simeq \fr{U(N)}{\bb{Z}_q}.
\end{equation}

Since the first homotopy group of the 
vacuum manifold is $\pi_1 (U(N)/\bb{Z}_q) = \bb{Z}$, 
there exist topologically stable vortices. 
Asymptotic behavior of vortex solutions far from the vortex core 
can be found as~\cite{Auzzi:2003fs}, 
\begin{equation}
( \vev{\phi_{cf}}_{\rm v} )\to 
\sr{\fr{\xi}{2}} \,{\rm diag}\,(1,...,1, e^{i\theta}), 
\end{equation}
\begin{equation}
 \vev{A_I}_{\rm v} \to
- \fr{1}{Nk+1} \,{\rm diag}\,(k, ..., k, k-(Nk+1))\der_I \theta, 
\end{equation}
\begin{equation}
 \vev{A_0}_{\rm v} = 0.
\end{equation}
Here,
the arrows indicate the limit $r \to \infty$,
and $\vev{...}_{\rm v}$ denotes the expectation value 
in the presence of a vortex, 
$\theta $ is the angle of the coordinate which is 
perpendicular to the vortex, $r$ is the distance 
from the vortex center, and 
$I = 1,2,3$ is the index of spatial coordinates.
For a finite distance from the core of the vortex, 
the configurations of the fields can be written by
\begin{equation}
( \vev{\phi_{cf}}_{\rm v} )= 
\sr{\fr{\xi}{2}} 
\mtx{
g(r)&  &  & 
\\
 &  \ddots &  &
\\
 &     & g(r) &
\\
 &   & &e^{i\theta} f(r)
}, \label{eq:vortex1}
\end{equation}
\begin{equation}
\begin{split}
\vev{A_I}_{\rm v}
=&
\Big(
-\fr{1}{Nk+1}
{\rm diag}\,
(k ,...,k, k-(Nk+1))
\\
&\hph{\Big(
\quad}
-\fr{1}{N(Nk+1)}
\bs{1}_N h^{U(1)}(r)
+\fr{1}{N}{\rm diag} \, (1,...,1,1-N) h^{SU(N)}(r)
\Big) \der_I \theta, 
\end{split}
\label{eq:vortex2}
\end{equation}
where the functions $f$, $g$, $h^{U(1)}$ and 
$h^{SU(N)}$ satisfy
\begin{equation}
 f(\infty) = g(\infty) =1, 
\quad
h^{U(1)} (\infty) = h^{SU(N)} (\infty) =  0,
\end{equation} 
and
\begin{equation}
 f(0) = 0, \quad g'(0) =0, 
\quad
h^{U(1)} (0) = h^{SU(N)} (0) = 1.
\end{equation} 
A Wilson loop $W$ 
across the plane perpendicular to the vortex strings
can be calculated as 
\begin{equation}
 \vev{W}_{\rm v}
= \vev{\tr {\cal P} \exp i\int_{\cal C} dx^I A_I  }_{\rm v}
=
N 
\exp\left(-2\pi i{k\over Nk+1}\right).
\label{eq:UN_wilson_vortex}
\end{equation}
Here, ${\cal C}$ is a circle at infinity which surrounds the 
vortex string, and 
${\cal P}$ denotes the path ordered product.
We observe that the unit of the Abelian magnetic flux 
is $1/N$ of the ANO magnetic flux,
whose configurations are given by
$\vev{\phi_{cf}}_{\rm v} \to \sr{\fr{\xi}{2}}e^{i\theta}\bs{1}_N$ and 
$\vev{A_I}_{\rm v} \to \fr{1}{Nk+1} \bs{1}_N \der_I \theta$ 
at the limit $r\to \infty$, and whose 
Wilson loop is 
$\vev{\tr {\cal P}\exp(i \int_{\cal C} dx^I A_I)}_{\rm v}  = 
N \exp(2\pi i\fr{1}{Nk+1})  = N \exp(-2\pi i\fr{Nk}{Nk+1})$.

The vortex configuration in Eqs.~(\ref{eq:vortex1}) 
and (\ref{eq:vortex2}) 
breaks the CFL symmetry $SU(N)_{{\rm c}+ {\rm f}}$ 
into subgroup $SU(N-1) \times U(1)$ around its core.
Consequently there appear Nambu-Goldstone modes
${\mathbb C}P^{N-1} \simeq SU(N)_{{\rm c}+ {\rm f}} /[SU(N-1) \times U(1)]$
localized around the vortex core, giving rise to the moduli 
${\mathbb C}P^{N-1}$.

\subsection{Dual $BF$-theory and topologically ordered phase}

Next,
we will show that there is an emergent $\bb{Z}_{Nk+1}$ 
two-form symmetry in addition to the $\bb{Z}_{Nk+1}$ one-form symmetry, and both of these symmetries are broken spontaneously
in the CFL phase.
In order to show them explicitly, it is convenient to 
dualize the effective action described by scalar fields 
to the action described by two-form gauge fields.
Here, we derive a dual topological action of the low-energy effective theory in \er{190117.0200}.
We consider the dynamics at lower energies compared to the mass of the amplitude fluctuation of $|\phi_{cf}|$, 
or the mass of the gauge fields. 
Since the vacuum manifold is $U(N)/\mathbb{Z}_{q}$,
one can always go to the gauge by the color rotation where 
the matrix $(\phi_{cf})$ is diagonalized:
\begin{equation}
 (\phi_{cf})
 = \sr{\fr{\xi}{2}}\, {\rm diag}\, (e^{i\chi_1},...,e^{i\chi_N}),
\label{190217.2214}
\end{equation}
where $\chi_i$ $(i=1,\cdots, N)$ are $2\pi$-periodic scalar fields. 
In this gauge, the low-energy action is given by%
\footnote{Such structure of the St\"uckelberg couplings 
between scalar fields and one-form fields are sometimes called 
an Abelian tensor hierarchy~\cite{deWit:2008gc,Hartong:2009az,Becker:2016xgv,Aoki:2016rfz,Yokokura:2016xcf}.}
\begin{equation}
\begin{split}
 {S}_{\rm NA, eff} 
&=  
\fr{\xi}{2}
\int \left|d \chi_1 - (k+1) a_1 - ka_2 - \cdots  - ka_N \right|^2
\\
&\quad
+\fr{\xi}{2}
\int \left|d \chi_2 - k a_1  - (k+1) a_2 -ka_3- \cdots  - ka_N \right|^2
\\ 
&\quad 
+\cdots
+\fr{\xi}{2}\int
\left|d \chi_N - k a_1 - \cdots  - ka_{N-1} - (k+1) a_N \right|^2.
\end{split}
\label{190120.1914}
\end{equation}
Here, $a_A$ $(A= 1,...,N)$  are the
one-form gauge fields $a_A = (a_A)_m \, dx^m$ which correspond to 
the Cartan's subalgebra of $U(N)_{\rm c}$.
We take the basis of the Cartan's subalgebra as 
\begin{equation}
 \begin{split}
& H_1  = \diag(1,0...,0), 
\quad
H_2 = \diag (0,1,0,...,0),
\, ...,\,
H_{N} = \diag (0,..,0, 1).
 \end{split}
\end{equation}
The gauge group of the action after the gauge in \er{190217.2214} is
\begin{equation}
U(1)_{H_1} \times \cdots \times U(1)_{H_{N}}.
\label{190129.0739}
\end{equation}

Following the steps in Sec.~\ref{sec:dual_BF_theory}, 
we obtain the dual action
\begin{equation}
 {S}_{\rm dual}
 = 
 \fr{1}{8\pi^2 \xi}
 \sum_i 
\int \left|db_i \right|^2
-
\frac{i}{2\pi}
K_{iA} 
 \int  b_i \wedge d a_A , 
\end{equation}
where $b_i$ are two-form gauge fields and the matrix $K_{iA}$ is given by 
\begin{equation}
 (K_{iA}) = {\bf 1}_N + k \bm{J}_N , 
 \label{eq:kmat}
\end{equation}
Here, $\bm{J}_N$ is the $N \times N$ matrix where every entry is $1$. 
Explicitly, $K_{iA}$ takes the form of
\begin{equation}
 (K_{iA}) =
\mtx{
k+1 & k &  \cdots  & k
\\
k & k+1 &   \cdots & k
\\
\vdots &  & \ddots  &\vdots
\\
k & k & \cdots  & k+1 
}. 
\end{equation}
The equation of motion varying $a_A$ gives $d b_i = 0$, since $K_{iA}$ has an inverse. 
So we can drop all the kinetic terms for $b_i$. 
This is in contrast with the case of 
$SU(3)$ gauge theories~\cite{Hirono:2018fjr}, 
where there remains a massless Nambu-Goldstone mode. 
Therefore, at a mass scale lower than that of the one-form fields 
and amplitude fluctuations, 
the effective action 
can be simply written as a $BF$-theory with matrix coupling, 
\begin{equation}
\begin{split}
 {S}_{U(N),BF}
& = 
-\fr{i}{2\pi} K_{iA} \int b_i \wed da_A.
\end{split}
\label{190117.1303}
\end{equation}
The gauge fields should satisfy the Dirac quantization condition, 
\begin{equation}
 \int_{\cal S} d a_A \in 2 \pi \bb{Z},
\quad
\int_{\cal  V} d b_i \in 2 \pi \bb{Z}, 
\end{equation}
where $\mathcal S$ and $\mathcal V$ are 2- and 3-dimensional submanifold 
without boundary. 
Incidentally, the same form of $K_i{}_A$ matrix (\ref{eq:kmat}) appeared 
in the description of the fractional Hall effect 
with filling factor $\nu = N / (Nk+1)$ by a Chern-Simons theory with matrix coupling~\cite{Zee:1991cf}.

Let us discuss the observables of the dual low-energy gauge theory (\ref{190117.1303}).
Similarly to the case of s-wave superconductors, 
the physically observable operators are the Wilson loops of the 
form\footnote{
The Wilson loop is not necessarily given by 
a dynamical particle. 
It can be thought of as a test particle with 
a possible charge (representation)~\cite{Aharony:2013hda}.
}, 
\begin{equation}
 W({\cal C}) = \sum_{A=1}^{N} \exp \(i\int_{\cal C} a_A\), 
\end{equation}
where $\mathcal C$ is a closed loop. 
As a remnant of the $U(N)$ gauge invariance, the physical Wilson lines should be invariant under the Weyl reflections $S_N$, and we here take an example of the fundamental Wilson line. 
There are also observable surface operators, 
\begin{equation}
 V_i ({\cal S}) = \exp\(i\int_{{\cal S}} b_i\), 
\end{equation}
where ${\cal S}$ is a 2-dimensional closed surface.
The set $\{V_1(\mathcal S),V_2(\mathcal S) ,\cdots, V_N(\mathcal S)  \}$ constitute the generators of all the physical surface operators. 
They are nothing but the non-Abelian vortices with minimal circulations when the surface $\mathcal S$ is extended in time and one spatial directions.

We can compute the correlation function between Wilson loops and vortex world-sheets as 
\begin{equation}
 \begin{split}
\vev{W({\cal C})V_i({\cal S})}
& = 
\sum_{A=1}^{N}\exp \( -  
2\pi i  (K^{-1})_{Ai} \link({\cal C},{\cal S})\)
\\
&= 
N \exp \(   
2\pi i  \fr{k}{Nk+1} \link({\cal C},{\cal S})\)
,
 \end{split}
\label{190119.1530}
\end{equation}
which reproduces the result of (\ref{eq:UN_wilson_vortex}). 
Here we used the fact that  the inverse of $K_i{}^A$ is given by 
$((K^{-1})_{Ai}) = \bm 1_N - \frac{k}{Nk+1} \bm J_N$. 
This relation shows that the theory has spontaneously broken 
$\mathbb{Z}_{Nk+1}$ one-form and dual two-form symmetries, 
thereby implying a topological order. 

\subsection{Adding theta term}

We have shown that the CFL phase 
of the $U(N)$ gauge theory with $N$-flavor Higgs fields is topologically ordered phase.
In four-dimensional gauge theories, we can further introduce the so-called 
theta term.
Here, we show that the background theta term gives rise to 
an effect on the correlation function as well as the vacuum expectation value (VEV) 
of the vortex surface 
operator.
First, we will see the effect of the theta term on the correlation function 
between the Wilson loop and the vortex surface operators.
Second, we will interpret the effect of the theta term on the 
correlation function as an anomaly between the periodicity 
of the theta term and the one-form symmetry~\cite{Gaiotto:2017yup, Tanizaki:2017bam, Kikuchi:2017pcp, Tanizaki:2018xto, Karasik:2019bxn, Cordova:2019jnf, Cordova:2019uob}.

\subsubsection{Deformation of correlation function}

Let us introduce a theta term as an external background field.
In the $U(N)$ gauge theory given by \er{190117.0200}, 
the theta term can be written as
\begin{equation}
 -\fr{i}{8\pi^2}\Theta \tr (F \wed F) -\fr{i}{8\pi^2}\Theta' (\tr F) \wed (\tr F),
\end{equation}
where $\Theta$ and $\Theta'$ are external fields.
We assume that $\Theta $ and $\Theta'$ have $2\pi$ periodicity.
In the following discussion, we consider only the $\Theta \tr{F \wed F}$ term 
for simplicity.
In the Abelian gauge, we assume that the gauge fields other than 
$a_A$ ($A=1,...,N$) are set to zero.
Under the condition, the theta term can be written as 
\begin{equation}
 -\fr{i}{8\pi^2} \Theta da_A \wed da_A.
\end{equation}

In order to see the effects of the theta term, 
we consider the dual $BF$-theory given by \er{190117.1303}.
Under the dual transformation, the theta term is not changed because 
the one-form gauge fields are not changed.
Thus, the $BF$-action with the theta term is given by
\begin{equation}
S_{U(N), BF, \Theta} = 
-\fr{i}{2\pi} K_{iA} \int b_i \wed da_A
- \fr{i}{8\pi^2}\int \Theta da_A \wed da_A. 
\end{equation}
Note that $BF$-theories with theta terms were considered in 
Refs.~\cite{Chan:2012nb,Jian:2014vfa,Lopes:2015sma,Chen:2015gma}.

Let us see the correlation function of the Wilson loop $W({\cal C})$ and the 
vortex surface operator $V_i ({\cal S})$ in the presence of the theta term.
The correlation function is given by 
\begin{equation}
 \vev{W({\cal C}) V_i({\cal S}) e^{\fr{i}{8\pi^2}\int \Theta da_A \wed da_A}}.
\end{equation} 
This correlation function can be evaluated as 
\begin{equation}
\begin{split}
&  \vev{W({\cal C}) V_i({\cal S}) e^{\fr{i}{8\pi^2}\int \Theta da_A \wed da_A}}
\\
 & 
= N \exp\( \fr{ 2\pi i k }{Nk+1} \link ({\cal C}, {\cal S}) \) 
\exp\(\fr{(2\pi)^2 i}{8 \pi^2} \(1- \fr{Nk^2 + 2k}{(Nk+1)^2}\) \int \Theta J_2({\cal S} ) \wed J_2({\cal S})\).
\end{split}
\label{191025.1633}
\end{equation}

Equation \eqref{191025.1633} 
shows that 
the linking phase is deformed if the surface ${\cal S}$ has a self-intersection number\footnote{For a step function like $\Theta$ that satisfies $d\Theta = J_3({\cal V})$ 
with a three-dimensional closed subspace ${\cal V}$,
the linking phase is deformed if the surface ${\cal S}$ has a self linking number on ${\cal V}$.}.
This effect of the theta term appears 
even if we set $W({\cal C})  =1$.
Therefore, the VEV of the vortex surface operator is deformed by the theta term.
Furthermore, the periodicity of the $\Theta$ is enlarged
from $2\pi$ to $ 2\pi (Nk+1)^2$ in the right-hand side of \er{191025.1633}.

\subsubsection{Anomaly between one-form symmetry and periodicity of $\Theta$}

In the previous section, we have explicitly shown that the correlation function is 
deformed by adding the theta term. 
However, the applicability of that computation is limited to the case $\xi \to \infty$, i.e., in the deep Higgs regime, so the details of the result may also be affected by finite $\xi$.
We here show that the interesting enlargement of $\Theta$-angle periodicity is topologically protected following the arguments in Refs.~\cite{Gaiotto:2017yup, Tanizaki:2017bam, Kikuchi:2017pcp, Tanizaki:2018xto, Karasik:2019bxn, Cordova:2019jnf, Cordova:2019uob}. 
This shows that the correlation function of extended objects must have the dependence on $\exp (i\Theta/(Nk+1)^2)$. 

To see it, we introduce the background gauge field $\mathcal{B}$ for $\mathbb{Z}_{Nk+1}$ one-form symmetry~\cite{Kapustin:2014gua}. This can be realized as the $U(1)$ two-form gauge fields with the constraint,
\begin{equation}
(Nk+1)\mathcal{B}=d \mathcal{C},
\end{equation}
where $\mathcal{C}$ is the $U(1)$ one-form gauge field. We postulate the invariance under the one-form gauge transformation, where the gauge parameter $\lambda$ is also the $U(1)$ one-form gauge field:
\begin{equation}
\mathcal{B}\mapsto \mathcal{B}+d \lambda, \; \mathcal{C}\mapsto \mathcal{C}+(Nk+1)\lambda. 
\end{equation}
Under this transformation, the dynamical $U(N)$ gauge field is transformed by
\begin{equation}
A\mapsto A+\lambda. 
\end{equation}
We can find the gauge-invariance of the scalar kinetic term by noticing that the following replacement of the covariant derivative,
\begin{equation}
D\phi\Rightarrow (d-i(k\,\tr[A]+A-\mathcal{C}))\phi,
\end{equation}
keeps the manifest one-form gauge invariance. We also have to replace the $U(N)$ field strength $F=dA+i A^2$ by 
\begin{equation}
F-\mathcal{B}. 
\end{equation}

Using this knowledge, we can now show that the periodicity of the theta angles is extended from $2\pi$ to $2\pi(Nk+1)^2$ for certain extended objects. To see it, let us compute the one-form gauge-invariant topological term as follows\footnote{Here, we only pay attention to $\Theta$, but the discussion for $\Theta'$ is also straightforward}: 
\begin{eqnarray}
{1\over 8\pi^2}\int \tr(F-\mathcal{B})^2&=&\underbrace{{1\over 8\pi^2}\tr F^2}_{\in\mathbb{Z}}-\underbrace{{1\over 4\pi^2}\int \tr[F]\wedge \mathcal{B}}_{\in {1\over Nk+1}\mathbb{Z}}+\underbrace{{N\over 8\pi^2}\int \mathcal{B}^2}_{\in {1\over (Nk+1)^2}\mathbb{Z}}. 
\end{eqnarray}
This shows that the periodicity of the partition function $Z[\mathcal{B},\Theta]$ with the background gauge field $\mathcal{B}$ is no longer $2\pi$ periodic. Indeed, we obtain 
\begin{equation}
Z[\mathcal{B},\Theta+2\pi (Nk+1)]=Z[\mathcal{B},\Theta]\exp i \left({N (Nk+1)\over 4\pi}\int \mathcal{B}^2\right). 
\end{equation}
Here, we shift the $\Theta$ by $2\pi(Nk+1)$ in order to eliminate the contribution from the mixed term, $\int \tr[F]\wedge \mathcal{B}$, and then the extra phase is determined only from the background gauge field. Since $\gcd(N, Nk+1)=1$, this expression proves the extension of $2\pi$ periodicity to $2\pi(Nk+1)^2$ periodicity. 

Before closing this section, let us make a few remarks.
The topologically ordered phase in the $U(N)$ gauge-Higgs system is very similar to that of an Abelian Higgs model with a charge $Nk+1$ Higgs field,
since both of the systems have spontaneously broken $\bb{Z}_{Nk+1}$ 
one- and two-form global symmetries.
It is therefore interesting if they are indeed the same. 

For the $U(N)$ gauge-Higgs system,
the numerical factor $1- \fr{Nk^2 + 2k}{(Nk+1)^2} $ in \er{191025.1633}
is determined by 
$\sum_A K^{-1}_{Ai}K^{-1}_{Ai} =1- \fr{Nk^2 + 2k}{(Nk+1)^2} $ ($i$ is not summed over), 
which is originated from the low-energy effective theory of the $U(N)$ gauge-Higgs system.
On the other hand,
for the Abelian Higgs model with the charge $Nk+1$ Higgs field, 
one can calculate the correlation function in the presence of the 
background theta term.
In this case, the numerical factor is 
$1-N \cdot\fr{Nk^2 + 2k}{(Nk+1)^2}= \fr{1}{(Nk+1)^2}$.
These numerical factors are different between those two theories, so this difference might be a candidate for distinction between topological phases of $U(N)$ gauge Higgs and Abelian-Higgs models. 

However, the anomaly discussed in this section is not the 't~Hooft anomaly in the usual sense, because the periodicity of theta angle is not symmetry. 
This kind of anomaly is sometimes called global inconsistency~\cite{Gaiotto:2017yup, Tanizaki:2017bam, Kikuchi:2017pcp, Tanizaki:2018xto, Karasik:2019bxn} or mixed anomaly with $(-1)$-form symmetry~\cite{Cordova:2019jnf, Cordova:2019uob}. 
The difference of global inconsistency leads to the fact that one of those two theories has to have a nontrivial (topological) order or those two theories must be distinguished as symmetry-protected topological orders. 
In our situation, both theories have nontrivial intrinsic topological orders, and the difference of above anomaly does not immediately mean the distinction as quantum phases. 
We therefore leave it as an open problem if the difference of the numerical factors in \er{191025.1633} give the distinction between topological orders of $U(N)$ gauge-Higgs and Abelian Higgs models.

\section{Summary and discussion}\label{sum}
In this paper, we have studied a $U(N)$ gauge theory with $N$-flavor 
scalar fields whose $U(1)$ charge is $Nk+1$ ($k\in \bb{Z}$). 
This theory has a $\mathbb{Z}_{Nk+1}$ one-form symmetry, and it is spontaneously broken in the CFL phase, which means that the phase is topologically ordered. 
The CFL phase hosts non-Abelian vortices appearing 
as topologically stable excitations, 
and the world sheets of these vortices are the generators of the $\mathbb{Z}_{Nk+1}$ one-form symmetry.
In order to see this, 
we have taken the Abelian dual of the low-energy effective description of the CFL phase,
and have obtained a $BF$-action. 
The Wilson loop operators as well as the surface operators are
described by the one-form and two-form gauge fields, respectively, in the 
$BF$-action. 
We have studied the braiding of the observable Wilson loops and surface operators, 
and found that they obey $\bb{Z}_{Nk+1}$ braiding statistics. 

We have discussed the deformation of the correlation function 
in the presence of the background theta term.
We have shown that the theta term gives rise to 
the effects on the correlation function as well as the VEV of the 
vortex surface operator.
We have further argued that the effect of the theta term on the 
correlation function can be understood as an anomaly between 
the periodicity of the theta term and the one-form symmetry.

The existence of topological order in the $U(N)$ gauge theory studied here 
is in contrast with the 
CFL phase of the $SU(N)$ gauge theory with $N$-flavor scalar fields, 
which is not topologically ordered \cite{Hirono:2018fjr}. 
In the latter case, there is an emergent two-form symmetry, 
but one-form symmetry is absent. 
And it turns out that the discrete two-form symmetry is the subgroup 
of a $U(1)$ two-form symmetry. 
Because a continuous two-form symmetry cannot be spontaneously broken in $(3+1)$ dimensions, the discrete two-form symmetry is always unbroken, 
hence there is no topological order. 
The crucial difference is that  $U(1)$ part is gauged in the current case,
and also the 
$U(1)$ charge of the scalar fields is taken to be a larger value, $Nk+1$.

There are several possible future directions.
One intriguing nature of non-Abelian vortices is that they have internal 
$\mathbb C P^{N-1}$ moduli inside them. 
The role of those modes in the topological properties of the system is to be investigated.
For instance, Yang-Mills instantons and magnetic monopoles 
are realized as sigma model instantons and kinks 
\cite{Shifman:2004dr,Hanany:2004ea,Eto:2004rz,Nitta:2010nd},
respectively in the $\bb{C}P^{N-1}$ model of the 
vortex world sheet.
In this paper, we have taken an Abelian duality after fixing the gauge
in \er{190217.2214}.
Instead of taking a particular gauge, 
it would be interesting to take a non-Abelian duality
in order to understand non-Abelian nature of the vortices~\cite{Seo:1979id,Hirono:2010gq}.
In particular, a coupling between non-Abelian two-form field and the vortex $\bb{C}P^{N-1}$ modes was derived in Ref.~\cite{Hirono:2010gq}.
Another direction is to examine the existence of topological order in a wider class of quantum field theories. 
Even if there is a topological order, coupling of the gauge fields to massless fermions might destroy the order.
When we add more flavors, 
vortices become non-Abelian semilocal vortices 
having non-normalizable size moduli 
\cite{Shifman:2006kd,Eto:2007yv}.
Since these vortices have polynomial tails of profile functions, 
interactions between vortices may destroy topological order. 
Although we have considered 
the $U(N)$ gauge group in this paper, 
more general gauge group of the type $[U(1) \times G]/C(G)$ 
would be possible \cite{Eto:2008yi}, where
$C(G)$ denotes the center of the group $G$.
Supersymmetric theories with topological vortices 
would allow us to do the analysis in a controlled way~\cite{Tong:2005un,Eto:2006pg,Shifman:2007ce,Shifman:2009zz}, 
for which the superfield formulation of duality of vortices 
in Ref.~\cite{Nitta:2018qqe} would be useful.

\subsection*{Acknowledgements}
The authors thank Masaru Hongo for useful discussions.
The work of M.~N.~and R.~Y.~is 
supported by the Ministry of Education,
Culture, Sports, Science (MEXT)-Supported Program for the Strategic Research Foundation at Private Universities `Topological Science' (Grant No.\ S1511006).
The work of M.~N.~is also supported in part by  
the Japan Society for the Promotion of Science
(JSPS) Grant-in-Aid for Scientific Research (KAKENHI Grant
No.~16H03984 and No.~18H01217).
The work of M.~N.~and R.~Y.~is 
also supported in part by a Grant-in-Aid for 
Scientific Research on Innovative Areas ``Topological Materials
Science'' (KAKENHI Grant No.~15H05855) 
and ``Discrete Geometric Analysis for Materials Design'' (KAKENHI Grant No.~17H06462) from the MEXT of Japan, respectively. 
The work of Y.~Hidaka is supported in part by Japan Society of Promotion of Science (JSPS) Grant-in-Aid for Scientific Research 
(KAKENHI Grants No.~16K17716, No.~17H06462, and No.~18H01211).
The work of Y.~Hidaka is supported  in part  by RIKEN iTHEMS Program.
The work of Y.~Hirono is supported in part by the
Korean Ministry of Education, Science and Technology,
Gyeongsangbuk-do and Pohang City for Independent
Junior Research Groups at the Asia Pacific Center for
Theoretical Physics. 
The work of Y.~T. was supported by RIKEN Special Postdoctoral Researchers Program until March 2019, and is supported by JSPS Overseas Research Fellowship. 

\appendix

\section{Notes on $BF$-theory and linking number}
\label{bflink}
Here, we review the derivations of \ers{190302.2148} 
and \eqref{190302.2001}.
First, we introduce delta function forms, intersection numbers, and 
linking numbers. 
Next, we show the derivations of
\ers{190302.2148} 
and \eqref{190302.2001} by using the delta function forms and 
linking numbers.

\subsection{Delta function forms}
For a $p$-dimensional subspace ${\cal C}_p$ 
in a $D$-dimensional space, we define 
the delta function $(D-p)$-form $J_{D-p}({\cal C}_p)$ as follows:
\begin{equation}
 \int_{{\cal C}_p} A_p  =
\int A_p \wed J_{D-p}({\cal C}_p).
\label{190302.2008}
\end{equation}
Here, $A_p$ is a $p$-form field.
In the flat space,
The delta function form can be explicitly written by
\begin{equation}
 J_{D-p}({\cal C}_p)
 = \fr{\epsilon_{m_1...m_p m_{p+1}...m_D}}{p!(D-p)!}
\(\int_{{\cal C}_p} \delta(x-y) dy^{m_1} \wed \cdots \wed dy^{m_p}\)
dx^{m_{p+1}} \wed \cdots \wed dx^{m_{D}}
\end{equation}
The exterior derivative on the delta function form 
is 
\begin{equation}
d J_{D-p} ({\cal C}_p)
=
(-1)^p  J_{D-(p-1)} (\der{\cal C}_p).
\label{190302.2103}
\end{equation}
Here, $\der$ denotes the boundary operator which satisfies 
$\der \der =0$.
The relation in \er{190302.2103} can be shown as follows:
\begin{equation}
\begin{split}
\int A_{p-1} \wed J_{D-(p-1)} (\der{\cal C}_p)
&=
\int_{\der{\cal C}_p} A_{p-1}
=
\int_{{\cal C}_p} d A_{p-1}  
=
\int d A_{p-1}  \wed J_{D-p} ({\cal C}_p) 
\\
&=
(-1)^{p} \int A_{p-1}  \wed d J_{D-p} ({\cal C}_p).
\end{split}
\end{equation}
Here, we have used 
$d(A_{p-1}\wed J_{D-p} ({\cal C}_p)) 
= dA_{p-1} \wed J_{D-p} ({\cal C}_p) + (-1)^{p-1} A_{p-1} \wed d
J_{D-p} ({\cal C}_p)$,
and used the fact that $J_{D-p} ({\cal C}_p) = 0$ at infinity.

\subsection{Intersection and linking number}
Let us consider $p$- and $q$-dimensional subspaces, ${\cal C}_p$ and ${\cal S}_q$. 
We denote the intersection of ${\cal C}_p$ and ${\cal S}_q$ as
${\cal I}_{p+q -D}$. 
The delta function form for ${\cal I}_{p+q -D}$ is given by
\begin{equation}
 J_{D -p + D -q}({\cal I}_{p+q -D}) 
=
 J_{D-p} ({\cal C}_p) \wed J_{D-q} ({\cal S}_q).
\end{equation}
If  $p + q = D$, the intersections of the subspaces are points. 
We can define the intersection number 
of ${\cal C}_p$ and ${\cal S}_q$ as
\begin{equation}
I ({\cal C}_p, {\cal S}_q)
= 
\int  J_{D} ({\cal I}_0)
=
\int  J_{D-p} ({\cal C}_p) \wed J_{p} ({\cal S}_q).
\end{equation}
If ${\cal C}_p$ has a boundary $\der {\cal C}_p$, 
the intersection number becomes the linking number
of $\der {\cal C}_p$ and ${\cal S}_q$:
\begin{equation}
 \link (\der {\cal C}_p, {\cal S}_q)
 = I({\cal C}_p, {\cal S}_q).
\end{equation}
\subsection{Correlation function in $BF$-theory and linking number}
Here, we briefly review the derivation of \er{190302.2001} in the 
path integral formulation following Ref.~\cite{Chen:2015gma}.%
\footnote{For a BRST invariant derivation, see Ref.~\cite{Oda:1989tq}.}

The correlation function given in \er{190302.2001} can be written as
\begin{equation}
\vev{W({\cal C}) V({\cal S})}
= {\cal N} \int {\cal D}A{\cal D} B
e^{\fr{ik}{2\pi}\int B \wed dA +i\int A\wed J_3({\cal C}) 
+ i\int B\wed J_2({\cal S})}.
\label{190118.1437}
\end{equation}
Here, ${\cal C}$ and ${\cal S}$ are 1- and 2-dimensional closed 
subspaces, respectively.
$J_3({\cal C})$ and $J_2({\cal S})$ are delta function forms 
defined in \er{190302.2008}.

In order to integrate \er{190118.1437}, 
we introduce the delta function forms whose exterior derivatives 
are $J_3({\cal C})$ and $J_2({\cal S})$ as in \er{190302.2103}.
In the 4-dimensional spacetime $\bb{R}^4$, 
there are 2- and 3-dimensional subspaces 
${\cal S}({\cal C})$ and ${\cal V}({\cal S})$ whose 
boundaries are ${\cal C}$ and ${\cal S}$:
\begin{equation}
 \der {\cal S}({\cal C}) 
= {\cal C}, \qquad \der {\cal V}({\cal S}) = {\cal S},
\end{equation}
respectively, 
since ${\cal C}$ and ${\cal S}$ are closed subspaces.
We can rewrite $J_3({\cal C})$ and $J_2({\cal S})$ by 
using exterior derivatives on $J_2({\cal S}({\cal C}))$ and $J_1 ({\cal V}({\cal S}))$: 
\begin{equation}
J_3({\cal C}) = J_3(\der {\cal S}({\cal C})) = dJ_2({\cal S}({\cal C})) , 
\quad
J_2({\cal S})
 = 
J_2 (\der {\cal V}({\cal S}) ) = -dJ_1 ({\cal V}({\cal S})),
\end{equation}
respectively.
Note that ${\cal S}({\cal C})$ and ${\cal V}({\cal S})$ are not unique:
one can add $\der {\cal V}$ and $\der \Omega$ to 
${\cal S}({\cal C})$ and ${\cal V}({\cal S})$, where 
${\cal V}$ and $\Omega$ are 3- and 4-dimensional subspaces 
in $\bb{R}^4$, respectively.
By using $J_2 ({\cal S}({\cal C}))$ and $J_1({\cal V}({\cal S}))$, 
we can show that $V({\cal S})$ acts on $W({\cal C})$ as 
a symmetry generator, and vice versa:
\begin{equation}
\begin{split}
&
 \vev{ V({\cal S})W({\cal C}) }
= 
{\cal N} \int {\cal D}A {\cal D}B \, 
e^{\fr{ik}{2\pi} \int B \wed d(A - \fr{2\pi}{k}J_1 ({\cal V}({\cal S})))
+ i \int A \wed J_3({\cal C})}
\\
&
=
{\cal N} \int {\cal D}A {\cal D}B \, 
e^{\fr{ik}{2\pi} \int B \wed dA
+ i \int A \wed J_3({\cal C}) 
+ \fr{2\pi i }{k} \int J_1({\cal V}({\cal S})) \wed J_3({\cal C})} 
= 
e^{- \fr{2\pi i}{k} \link ({\cal C},{\cal S})}
\vev{W({\cal C})}, 
\end{split}
\label{190302.2149}
\end{equation}
and 
\begin{equation}
\begin{split}
& \vev{V({\cal S})W({\cal C})}
= 
{\cal N} \int {\cal D}A {\cal D}B \, 
e^{\fr{ik}{2\pi} \int (B + \fr{2\pi}{k}J_2 ({\cal S}({\cal C}))) \wed dA 
+ i \int B\wed J_2 ({\cal S} )}
\\
&
=
{\cal N} \int {\cal D}A {\cal D}B \, 
e^{\fr{ik}{2\pi} \int B \wed dA 
+ i \int B \wed J_2 ({\cal S} ) - \fr{2\pi i}{k}\int
 J_2 ({\cal S}({\cal C})) \wed J_2 ({\cal S} ) }
= 
e^{- \fr{2\pi i}{k} \link ({\cal C},{\cal S})}
\vev{V({\cal S})},
\end{split}
\end{equation}
where we have used the reparametrizations 
 $A \to A + \fr{2\pi}{k} J_1({\cal V}({\cal S}))$ and 
$B \to B - \fr{2\pi}{k} J_2({\cal S}({\cal C})) $, 
respectively.
We can similarly show $\vev{V({\cal S})} =1$ 
in \er{190302.2148} and $\vev{W({\cal C})}=1 $ as follows:
\begin{equation}
\vev{V({\cal S})} 
=
{\cal N} \int {\cal D}A {\cal D}B \, 
e^{\fr{ik}{2\pi} \int B \wed d(A - \fr{2\pi}{k}J_1 ({\cal V}({\cal S})))}
=
{\cal N} \int {\cal D}A {\cal D}B \, 
e^{\fr{ik}{2\pi} \int B \wed dA}
 = 1,
\end{equation}
and
\begin{equation}
\begin{split}
 \vev{W({\cal C})} 
&=
 {\cal N} \int {\cal D}A{\cal D} B
 e^{\fr{ik}{2\pi}\int (B + \fr{2\pi}{k} J_2({\cal S}({\cal C})))
\wed dA }
=
 {\cal N} \int {\cal D}A{\cal D} B
 e^{\fr{ik}{2\pi}\int B \wed dA }
=
 1.
\end{split}
\label{190302.2209}
\end{equation}
Note that \ers{190302.2149} and \eqref{190302.2209} show 
that the correlation function in \er{190118.1437} gives 
us the linking number of ${\cal C}$ and ${\cal S}$:
\begin{equation}
\vev{W({\cal C})V({\cal S})} = e^{- \fr{2\pi i}{k} \link ({\cal C},{\cal S})}. 
\end{equation}

\providecommand{\href}[2]{#2}
\end{document}